\documentclass{article}

\PassOptionsToPackage{numbers, sort&compress}{natbib}

\usepackage[preprint]{neurips_2023_ml4ps}
\usepackage{graphicx} 
\usepackage[utf8]{inputenc} 
\usepackage[T1]{fontenc}    
\usepackage{hyperref}       
\usepackage{url}            
\usepackage{booktabs}       
\usepackage{amsfonts}       
\usepackage{nicefrac}       
\usepackage{microtype}      
\usepackage{xcolor}         
\usepackage{caption}        
\usepackage{amsmath}        
\newcommand{\vect}[1]{\boldsymbol{#1}}

\title{GalacticFlow: Learning a Generalized Representation of Galaxies with Normalizing Flows}
\author{%
  Luca Wolf\\
  Interdisciplinary Center\\ for Scientific Computing,\\
  University of Heidelberg,\\
  Im Neuenheimer Feld 205,\\ D-69120 Heidelberg\\
  \texttt{luca.wolf@stud.uni-heidelberg.de} \\
    \And
     Tobias Buck\\ 
  Interdisciplinary Center\\ for Scientific Computing,\\  University of Heidelberg,\\ Im Neuenheimer Feld 205,\\ D-69120 Heidelberg\\
  \texttt{tobias.buck@iwr.uni-heidelberg.de} \\
}

\begin{document}

\maketitle
\begin{abstract}
    State-of-the-art galaxy formation simulations generate data within weeks or months. Their results consist of a random sub-sample of possible galaxies with a fixed number of stars.
    We propose a ML based method, \texttt{GalacticFlow}, that generalizes such results. We use normalizing flows to learn the extended distribution function of galaxies conditioned on global galactic parameters. \texttt{GalacticFlow} then provides a continuized and condensed representation of the ensemble of galaxies in the data. Thus, essentially compressing large amounts of explicit simulation data into a small implicit generative model. Our model is able to evaluate any galaxy eDF given by a set of global parameters and allows generating arbitrarily many stars from it. We show that we can learn such a representation, embodying the entire mass range from dwarf to Milky Way mass, from only 90 galaxies in $\sim18$ hours on a single RTX 2080Ti and generate a new galaxy of one million stars within a few seconds.
\end{abstract}

\section{Motivation}

In astrophysics, distribution functions (DFs) play a crucial role in modeling the properties and behavior of galaxies. DFs provide a mathematical description to characterize the statistical properties of the stars in a galaxy. In galactic dynamics, the DF, often denoted as $f(\vect{x}, \vect{v})$, describes the probability of finding a star with specific phase space coordinates within the galaxy \citep{Binney2010}. 
They have been widely used to describe the dynamics of collisionless systems such as galaxies via Boltzmann's equation \citep{galaxydynamics2008} and allow to calculate various properties, such as the mass, velocity dispersion, and density profile via integration of the DF over different regions of phase space.
Since galaxies typically consist of various populations of stars with different properties, such as age or chemical composition (metallicity) but similar orbit distributions the concept of extended DFs has been introduced \citep{Sanders2015}. Those eDFs combine the phase space distribution with additional stellar properties in order to model their joint distribution and allow for more complex relations among stars from different populations.
In general, those eDFs can become quite complex especially for galaxies with a wide range of stellar populations and complex dynamical histories such that they become analytically intractable.
Here we draw on the fact that ML concepts such as normalizing Flows (NFs) are able to describe such complex DFs. Furthermore, automatic differentiation allows to easily compute gradients of the approximated DFs enabling to feed the neural DFs into existing mathematical modelling frameworks such as Boltzmann's equation.  
Normalizing\section{Related Work: Normalizing flows in astrophysics}
\begin{figure}
\begin{center}
    \includegraphics[width=.7\textwidth]{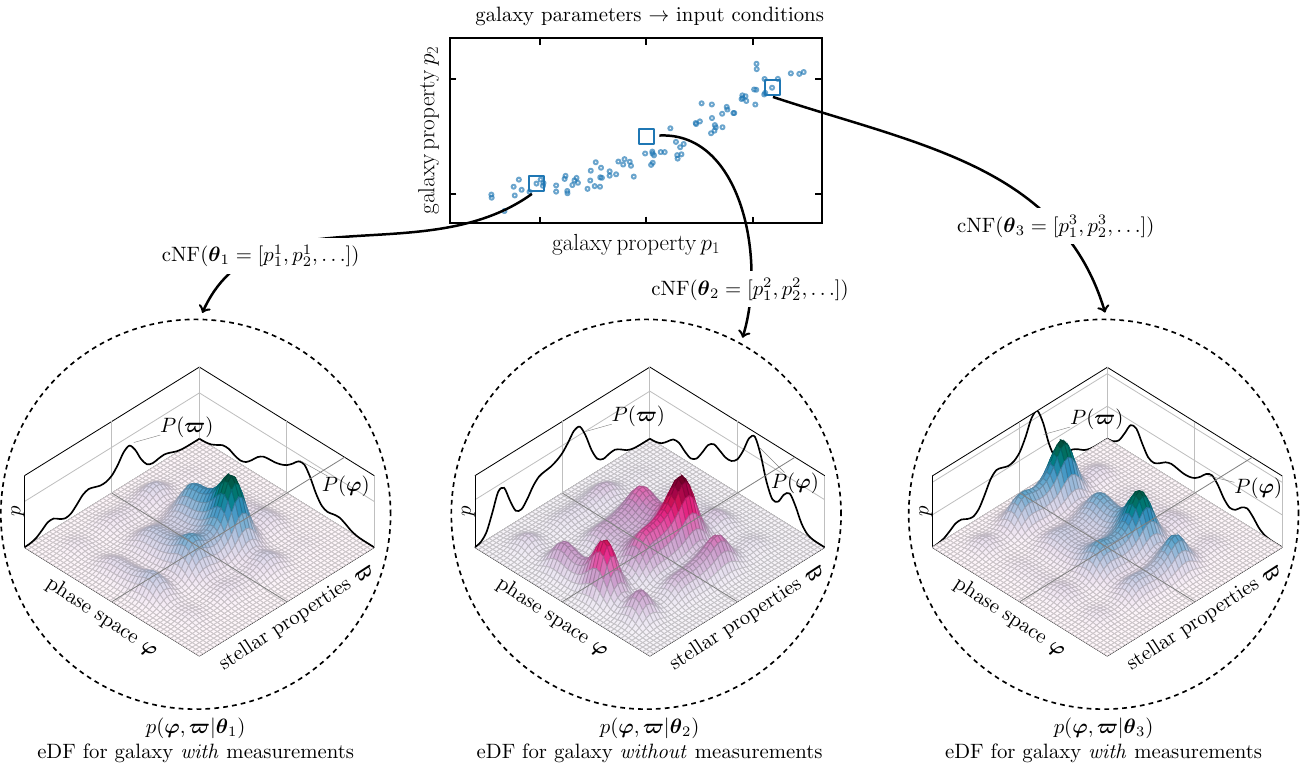}
    \caption{Overview of the GalacticFlow network. We use global galaxy properties $p_i$ as input conditions $\vect{\theta}$ to the conditional normalizing flow (cNF), $p(\vect{\varphi},\vect{\varpi}|\vect{\theta})$, to generate extended distribution functions (eDFs) $p_{\vect{\theta}}(\vect{\varphi},\vect{\varpi})$ of galaxies. Where $\vect{\varphi}$ denotes the phase space coordinates and $\vect{\varpi}$ the additional stellar quantities like age and metallicity.} 
    \label{fig:scheme}
\end{center}
\end{figure}

\enlargethispage{\baselineskip}
NFs are a machine learning (ML) tool, that allows to learn complex, high-dimensional, multi-modal probability distribution functions (pdf) from given data points $\vect{x}_i \in \mathbb{R}^n$.
Besides their study in probabilistic ML and application to many general tasks, such as image generation, they have recently been found useful in a wide range of physics problems. 
Particularly in astrophysics, they are used to model the phase space density of stars and infer the Galactic dark matter density and gravitational potential via Boltzmann's equation \cite{deep_potential2,pdf_DM}. \citep{Cranmer2019} applied NFs to describe the color-magnitude diagram of the Gaia data, and \citep{Reiman2020} used conditional NF models (cNFs) to model quasar continua while \cite{elements_matter} used them to model the element distribution of stars.
Here we go one step further and not only learn the 6d phase space density of stars, but their 10d eDF that further covers age, metallicity, and oxygen/iron abundance. We describe galactic DFs via NFs that are trained on state-of-the-art simulations of galaxy formation to directly learn the stellar eDFs from high quality data. In particular, we use cNFs to describe the eDFs conditioned on global galaxy properties such as total stellar mass, average meatallicity or star formation history, that provide additional contextual information when sampling. As such, we describe a population of galactic DFs with a single cNF model, illustrated in Fig.~\ref{fig:scheme}. 
The idea behind \texttt{GalacticFlow} is based on the fact that galaxy morphology is linked to underlying physical parameters via various scaling relations, most famously, the Mass-Metallicity  \citep{Tremonti2004,Gallazzi2005}, Tully-Fisher \citep{Tully1977} and Faber-Jackson relation \citep{Faber1976}. Furthermore, galaxies grow hierarchically, i.e. by merging with others, resulting in older galaxies being heavier and more elliptical, while younger galaxies are lighter and more spiral-like giving rise to the famous Hubble sequence \cite{Hubble1926}.
Learning these different DFs with a NF promises a compressed model, that allows generating arbitrarily many stars from exact DFs of arbitrary galaxies in very short times. In comparison to other generative models, this further allows direct evaluations of those DFs.

\section{Method}
\enlargethispage{\baselineskip}
A normalizing flow constructs a diffeomorphism $f^{-1}$, such that $\vect{y}_i=f^{-1}\left(\vect{x}_i\right)$ follows a simple base distribution \citep[e.g. a gaussian][]{Kobzev2019,NFs_KL}. 
The change of variable theorem relates $f^{-1}$ to the data's modeled pdf via:
    $\log p\left(x\right)=\log p'\left( f^{-1}\left(x\right)\right) + \log \left|J_{f^{-1}}\left(x\right)\right|$,
where $p'$ is the prior distribution in $y$-space. 
To find $f^{-1}$, that yields Gaussian distributed $y$-values, one sets $p'=\mathcal{N}(0,1)$ and maximizes $\mathcal L=\overline{\log p \left( x\right)}$, the mean log likelihood of the data. This is equivalent to minimizing the Kullback–Leibler divergence of $p_\mathrm{true}$ from $p$ \citep{NFs_KL}.
After training, we obtain an exact modeled pdf and are able to generate arbitrarily large samples by drawing $y \sim \mathcal{N}(0,1)$ and applying $f$.
$f^{-1}$ is composed of several functions $f^{-1}_1,\dots,f^{-1}_m$ that use the chain rule to compute $\log p\left(x\right)$.
These functions can be chosen as simple parametric functions, e.g. splines, with neural networks (NNs) determining their parameters individually per data point. 
Coupling layers \citep[e.g.][]{nice, realnvp}, maintain the first half of dimensions and input them into NNs for the transformation's parameters  of the second half. This is followed by a permutation of the dimensions and repeated several times and satisfies the requirements of a diffeomorphism with efficiently computeable inverse and Jacobian determinant.
\subsection{Model architecture}
Our NF architecture uses very flexible rational-quadratic spline coupling layers \cite{neural_spline_flows} alternating with invertible $1\times1$ convolutions as generalized permutations \cite{glow} as presented by \cite{neural_spline_flows}. We also adapt "double coupling layers", as used by \cite{elements_matter} which transforms the second half of dimensions with parameters determined from the first half followed by a transformation of the first half with the new second half. This choice enables both forward ($f^{-1}$) and backward ($f$) transformations in a single pass through the NNs, contrasting with auto-regressive methods. This allows training and sampling to be fast at the same time.
This base NF can easily be expanded to learn conditional pdfs. To achieve this, we follow \cite{elements_matter} and input conditional values into a second NN and multiply the outputs of both networks to determine the transformation parameters. These conditional values must be provided for every point in the training set and as necessary additional information for each sample drawn. 
We choose 14 flow layers (convolution layer followed by coupling layer), 
$K=10$ spline bins per coupling layer and 3 hidden layers with each 128 neurons for the coupling layer NNs. In our tests we found this architecture to be working best, as smaller models struggled to fit the data while larger models were over fitting. However, we did not perform extensive cross-testing of all hyper-parameters. Our final cNF model operates on 10D data and employs a 4D condition space that we describe further below. 
With this architecture the model has about $3\cdot10^6$ trainable parameters, taking about $11\mathrm{MB}$ of disk space. 
\begin{figure}
\begin{center}
    \includegraphics[width=.7\textwidth]{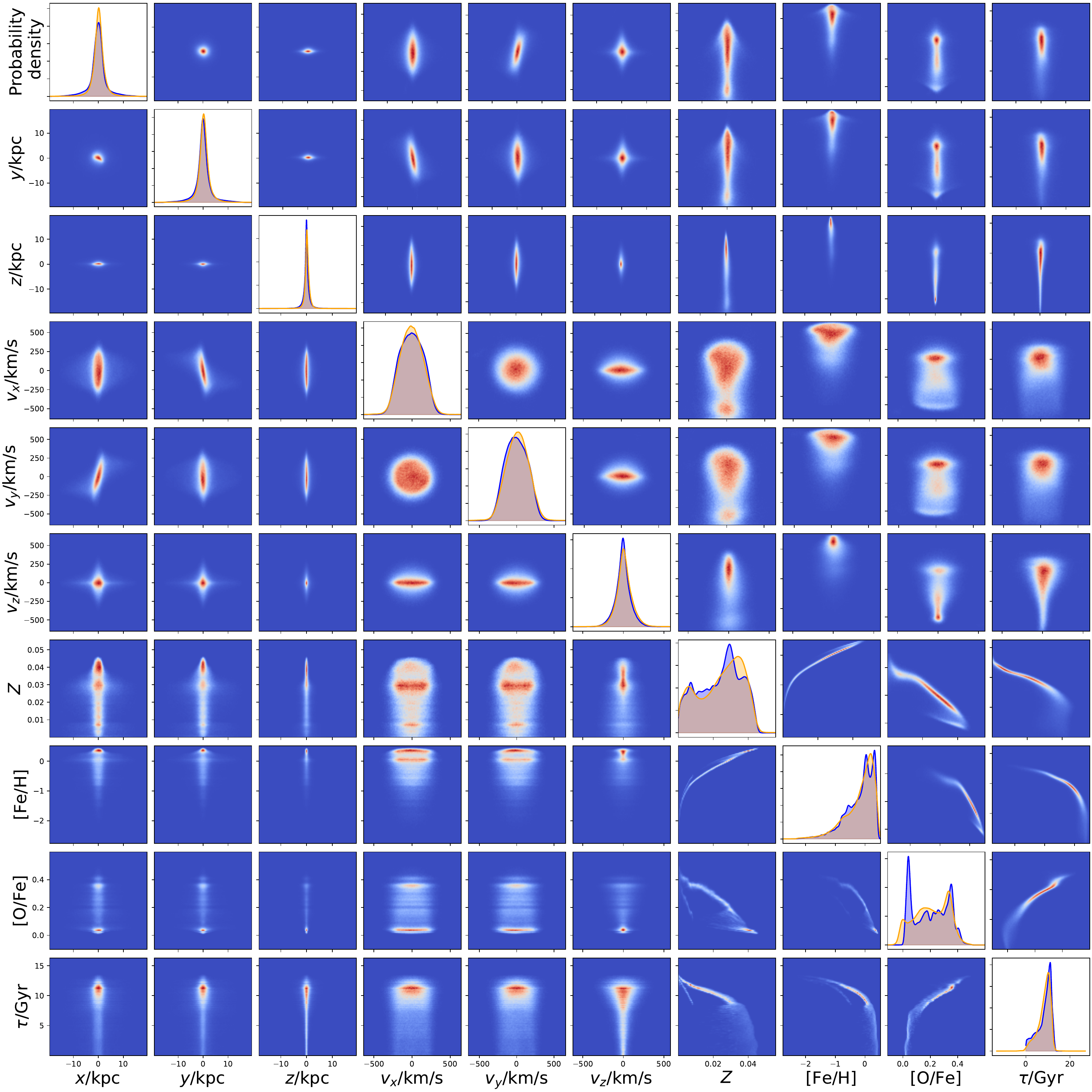}
    \caption{Corner plot of the 10D stellar parameter space. The diagonal shows a Gaussian KDE of the respective marginal distribution with data from the validation set in blue and model in orange. The off-diagonal shows 2D histograms of ground truth data in the bottom left corner and the corresponding \texttt{GalacticFlow} model the top right corner. Note that the images for the samples we have flipped x and y axis to have the same orientation as the data, for better comparison. They then share the same axis and the same histogram grid with the respective data plot.} 
    \label{fig:prediction}
\end{center}
\end{figure}
\enlargethispage{\baselineskip}
\subsection{Training}
To demonstrate our model, we train a \texttt{GalacticFlow} model on simulation data from the NIHAO project \cite{NIHAO_1, NIHAO_UHD, Buck2019, Buck2020} that models 95 galaxies from dwarf galaxy to the Milky Way mass scale. From this simulation suite we extract for each galaxy the 10 dimensional joint distribution of stellar position $\vect{x}$, velocity $\vect{v}$, total metallicity $Z$, iron and oxygen abundance [Fe/H], [O/Fe] and stellar age $\tau$. Additionally, we record easy to observe galactic properties such as total stellar mass $M_{\star}$, median stellar age $\tau_{50}$, the 10-th percentile stellar age $\tau_{10}$ and the average stellar metallicity $Z$ for each galaxy to be used as input conditions to the cNF.
We normalize the data to zero mean and unity standard deviation. Total stellar mass is transformed to log space before normalizing, for a more natural spacing. The inverse of this transformation needs to be applied after sampling and the Jacobian needs to be respected in case of pdf evaluation.
For training we use the RAdam optimizer \cite{RAdam} and an exponential learning rate schedule with an initial learning rate of $\textrm{lr}=9\cdot 10^{-5}$ and a decrease of a factor $\gamma = 0.998$ every 10 and once the learning rate falls below $3\cdot10^{-6}$, every 120 optimisation steps. 
We train 20 different models, where in each model 5 randomly chosen galaxies are left out from the training data and remain reserved for validation only.
Using a batch size of 1024, we train for 10 epochs on a single RTX 2080Ti, which takes about 18 hours of wall clock time. Drawing a sample from this flow takes about $3.5\mathrm s$ per GPU per million stars, in the limit of large sample sizes.
\section{GalacticFlow: Learning a generalized generative model for galaxies}
\begin{minipage}{0.5\textwidth}
We evaluate the the quality of the \texttt{GalacticFlow} model by comparing 100 generated galaxy samples with their ground truth data from the validation set.
In this way we can judge how the model performs in providing a generalized representation that interpolates in galactic parameter space. A "generated galaxy" in this context refers to a sample from the flow with conditions matching those of a ground truth galaxy, including the same number of stars.
It is vital to note, that we neither expect nor intend a perfect agreement between a \texttt{GalacticFlow} model and a real galaxy. After all, each galaxy possesses a unique merger history and distinctive characteristics that cannot cannot be generalized with only 4 conditions. Thus, our objective with \texttt{GalacticFlow} is to capture general trends in galactic eDFs rather than memorizing individual formation histories. An overly close agreement would only be possible on an over-fitted training galaxy. In Fig.~\ref{fig:prediction} we show a qualitative
\end{minipage}
\begin{minipage}{0.05\textwidth}
\end{minipage}
\begin{minipage}{0.48\textwidth}
    \includegraphics[width=1.025\textwidth]{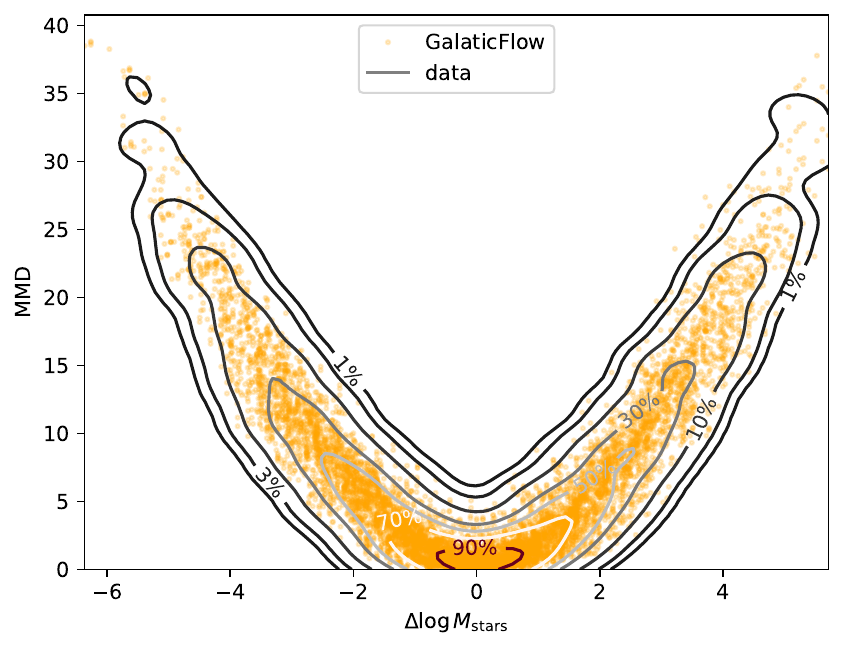}
    \captionof{figure}{
    Linear maximum mean discrepancy vs. difference in log stellar mass $\Delta \log M_{\rm{stars}}$ between \texttt{GalacticFLow} models and data galaxies (orange points) and corresponding ground truth galaxies and data galaxies (contours).}
    \label{fig:mmd_vs_M}
\end{minipage}
 comparison of one example galaxy from the validation set with its \texttt{GalacticFlow} model in a corner plot. 
The agreement of model and data can be deemed quite good. \texttt{GalacticFlow} demonstrates the capacity to interpolate and generate realistic galaxies for parameter sets not seen during training. 
Across the entire mass range,  it successfully captures various shapes of stellar eDFs. 
However, some challenges are evident in reproducing sharp peaks, which may relate to the naive sampling procedure \cite{nf_sails}. 
We anticipate that model quality and its ability to recognize trends will improve considerably with a more diverse training set. 
Finding a metric to quantitatively evaluate the performance of our model is challenging, as we need a sample-based difference measure between two 10D distributions. In the spirit of our model we also require the metric to measure 'physical closeness', at least in a sense that, e.g., galaxies of similar mass are deemed close. We choose to employ linear maximum mean discrepancy \cite[MMD][]{mmd} as we found it to fulfill these requests. 
We generate a sample of an unseen galaxy and compute the linear MMD to all 90 data galaxies, recording the difference in log stellar mass $\Delta \log M_{\rm{stars}}$ between model and data galaxy. We repeat this process for all 5 unseen galaxies of all 20 models, to calculate a relation of statistical distance as a function of $\Delta \log M_{\rm{stars}}$ as shown in Fig.~\ref{fig:mmd_vs_M}. This relation enables us to evaluate the quality of generalization and the recognition of general, known trends by the model. We show here MMD vs. $\Delta \log M_{\rm{stars}}$ as stellar mass is a primary condition and a key predictor of galaxy morphology. The clear trend of increasing MMD with increasing magnitude of $\Delta \log M_{\rm{stars}}$ and the statistical similarity between model and data (points vs. contours) indicates that \texttt{GalacticFlow} eDFs align well with those of real galaxies of similar mass.
Thus, the model effectively learns a meaningful generalization of galactic parameter space.
\enlargethispage{\baselineskip}
\section{Conclusion}
We find that \texttt{GalacticFlow} can learn a generalized representation of galaxies with conditional normalizing flows, that captures and respects the underlying galactic parameter space very well and interpolates it nicely. After a moderate training time ($\sim18$h), we obtain a compact generative model ($3\cdot10^6$ trainable parameters, $\sim11\mathrm{MB}$), that allows to generate realistic galaxies in any desired size for any desired set of galactic parameters in short times ($3.5\mathrm s$ per GPU per million stars). 

Our code is publicly available on GitHub\footnote{\url{https://github.com/luwo9/GalacticFlow}} and the galaxy data can be found on Zenodo\footnote{\url{https://doi.org/10.5281/zenodo.8389554}}. This includes our algorithms for pre-processing the data as well as the source code for the cNF. The code is documented and can easily adapted for reuse. Furthermore we provide a user friendly API for loading our pre-trained models, training your own models and generating/sampling galaxies as desired.

\section*{Broader impact statement}

The authors are not aware of any immediate ethical or societal implications of this work. This work purely aims to aid scientific modelling and proposes to apply normalizing flows to learn the distribution function of stars in galaxies. Thinking more broadly, the concept of  \texttt{GalacticFlow} is to extract and interpolate high-dimensional data from spares simulation results to facilitate faster and more versatile modelling. This might help to recognize how much information is already contained in legacy data that can be made accessible for better modelling and as such help to save energy by running less expensive simulations.  

\bibliography{bibliography}
\end{document}